\documentclass[structabstract,letter,usenatbib]{aa}  
\usepackage{graphicx,color}
\usepackage{ifpdf}
\ifpdf
  \DeclareGraphicsExtensions{.pdf}
\else
  \DeclareGraphicsExtensions{.eps}
\fi
\usepackage{txfonts}
\usepackage{natbib}
\bibpunct{(}{)}{;}{a}{}{,} 
\newcommand{\Msun}{\mbox{\,$\rm M_{\odot}$}}
\newcommand{\Lsun}{\mbox{\,$\rm L_{\odot}$}}
 
\newcommand{\teff}{$T_{\rm eff}$}

\newcommand{\logg}{$\log g$}
\begin{document}

\title{A newly discovered stellar type: dusty post-red giant
  branch stars in the Magellanic Clouds.}
\titlerunning{Dusty post-RGB stars}
\author{D. Kamath\inst{1}, P. R. Wood\inst{2}, H. Van Winckel\inst{1},
J.D. Nie\inst{3}}
\institute{Instituut voor Sterrenkunde, K.U.Leuven, Celestijnenlaan 200D bus 
2401, B-3001 Leuven, Belgium \and  Research School of Astronomy and Astrophysics, Mount Stromlo 
Observatory, Weston Creek, ACT 2611, Australia \and Key Laboratory of Optical Astronomy, National Astronomical Observatories, 
Chinese Academy of Sciences, Beijing 100012, China}

\offprints{Devika.Kamath@ster.kuleuven.be}
\date{}
\authorrunning{D. Kamath et al.}
\titlerunning{Dusty Post-RGB stars in the Magellanic Clouds}

\abstract {We present a newly discovered class of low-luminosity,
  dusty, evolved objects in the Magellanic Clouds. These objects have dust excesses, stellar
parameters,  and spectral energy distributions similar to those of dusty 
post-asymptotic giant branch (post-AGB) stars. However, they have
lower luminosities and hence lower masses. We suggest that they have evolved off the red
giant branch (RGB) instead of the AGB as a result of binary
interaction.} {In this study we aim to place these objects in an
evolutionary context and establish an evolutionary
connection between RGB binaries (such as the sequence E variables) and
our new sample of objects.} 
{ We compared the theoretically predicted birthrates of the progeny of RGB
  binaries to the observational birthrates of the new sample of
  objects.} 
{We find that there is order-of-magnitude agreement between the observed and 
predicted birthrates of post-RGB stars.  The sources of uncertainty in the birthrates
are discussed; the most important sources are probably the observational incompleteness factor and the
post-RGB evolution rates.  We also note that mergers are relatively common 
low on the RGB and that stars low on the RGB with 
mid-IR excesses may recently have undergone a merger.}
{Our sample of dusty post-RGB stars most likely provides the first observational
evidence for a newly discovered phase in binary evolution: post-RGB binaries with circumstellar dust.}

\keywords{Stars: binaries: general --Stars: evolution -- (galaxies:) Magellanic Clouds --
Methods: observational -- Methods: numerical}

\maketitle
\section{Introduction}
\label{intro}
\vspace{-0.75em}
The evolutionary fate of a star on the giant branches (the red
giant branch [RGB] and the asymptotic giant branch [AGB])
significantly depends on whether the star evolves via a single or a
binary evolutionary channel.

The nuclear-burning lives of single low- to intermediate-mass stars are terminated during the AGB phase when a super-wind with
mass-loss rates of up to 10$^{-4}$\,\Msun\,yr$^{-1}$ reduces the hydrogen
envelope to low values \citep[$\la$\,10$^{-2}$\,\Msun,
see][]{schoenberner83,vw93}.  Subsequently, within $\sim$\,10$^{2}$\,$-$\,10$^{4}$ years, the
star evolves to higher temperatures through the post-AGB phase (defined
approximately as the temperature range from the AGB to 10$^{4}$\,K)
with an almost constant luminosity. Single post-AGB stars are
surrounded by the expanding dusty matter that is expelled by the stellar wind. The
stellar photospheric emission is absorbed by the dust in the wind and
is re-emitted, leading to stars with high mid-IR excess. See \citet{vanwinckel03}
for a review of the post-AGB phase.  Beyond the post-AGB phase, 
when \teff~$\ga$ 2\,$\times$\,10$^{4}$\,K, the
star passes through the planetary nebula (PN) and white dwarf (WD) phases.

For low- to intermediate-mass stars in binary systems a different
mechanism can terminate AGB evolution. The large expansion that occurs when the star
is on the AGB can cause the primary star to overflow its Roche lobe. The
evolutionary outcome depends primarily on the time at which the star fills its
Roche lobe. A star that reaches the
tip of the AGB without filling its Roche lobe will evolve as a
single star does, resulting in the formation of a PN central star on a wide orbit \citep{moe06}. 
For a star that fills its Roche lobe on the AGB, a common envelope (CE) event usually occurs, which
results in either a close binary or a stellar merger
\citep{ivanova13}. Alternatively, for a small range of mass ratios, stable mass transfer via Roche lobe overflow
can occur, leading to the formation of an intermediate-period
binary. Post-AGB stars surrounded by
circumbinary discs are most likely examples of such systems \citep{vanwinckel09}.

Stars in binary systems not only interact while on the AGB, but can
also interact while on the RGB, which results in large amounts of mass loss,
followed by evolution off the RGB
\citep[e.g.][]{han02,nie12}. Here, the outcome of a CE event
is thought to be either a close binary containing a
low-mass He-core white dwarf or a core-He burning
subdwarf B star \citep{webbink84,heber09}, or a stellar merger in which the secondary star 
merges with the red
giant envelope.  The merged star will evolve further as a
more massive single star, initially with a dusty circumstellar envelope
left over from the merger process. Roche-lobe filling can also lead to RGB
termination by stable mass transfer for a small range of mass ratios.
In this case, subsequent evolution beyond the RGB is to higher \teff~values
at near-constant luminosity \citep{driebe98}, initially through the post-RGB phase, which is defined
approximately as the temperature range from the RGB to \teff\,$\approx$ 10$^{4}$\,K.

All of the above-mentioned theoretical scenarios agree well with those
borne out by population-synthesis models of \citet{han02} and \citet{nie12}. Although these evolution schemes seem
straightforward, there is little understanding from first principles
of the different important physical processes that govern these 
binary interaction processes. To fully understand single and binary stellar
evolution, observations of systems that have just evolved off the
AGB and RGB are essential.

In this letter we report on the first observational evidence for
objects that appear to be dusty post-RGB systems.  We aim to place these
objects into the context of binary evolution and compare the observed
sample to the predictions of
population synthesis models that simulate the fate of binary red
giants. 
\vspace{-1.5em}
\section{Low-luminosity, dusty post-RGB sample} 
\label{data}
\vspace{-0.75em}
We have carried out an extensive search for
optically bright dusty post-AGB 
candidates in the Magellanic Clouds. We performed an extensive
low-resolution optical spectral survey with the AAOmega multi-fibre
spectrograph mounted on the Anglo Australian telescope, which resulted
in a clean sample of well-characterised objects
with spectroscopically determined stellar parameters (\teff, \logg,
[Fe/H] and E[$B-V$]) spanning a wide range in luminosities in the
Small Magellanic Cloud [SMC] \citep[][Paper I]{kamath14} 
and the Large Magellanic Cloud [LMC] \citep[][Paper II]{kamath15}.

\begin{figure}
\begin{center}
\resizebox{0.85\hsize}{!}{ \includegraphics{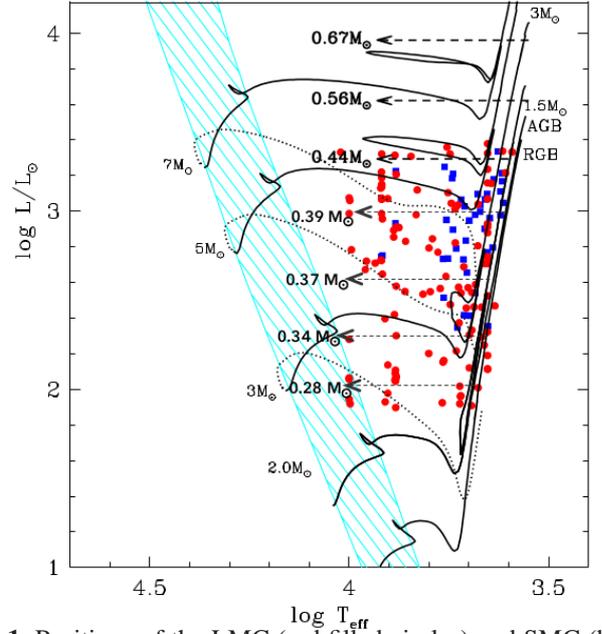}}
\vspace{-1.0em}
\caption{Positions of the LMC (red filled circles) and SMC (blue
  filled squares) post-RGB candidates on the HR diagram. The main-sequence is represented as
  a cyan and grey cross-hatched region. The black solid lines represent
  evolutionary tracks starting from the main sequence and continuing
  up to the AGB-tip with an initial composition Y=0.23, Z=0.004 
  \citep{bertelli08,bertelli09}. The black dotted lines
  represent PMS evolutionary tracks for an initial composition Y=0.238, Z=0.004 
  \citep{tognelli11}. The masses of the evolutionary
  tracks are indicated in the plots, with the pre-main-sequence and main-sequence masses
  shown on the left side of the main-sequence band and the RGB-tip masses on  the right side of the plots. The black dashed arrows schematically
  represent the post-AGB and post-RGB evolutionary tracks (the masses
  of the post-AGB and post-RGB stars are given at the arrow ends). The masses
  for the post-AGB evolutionary tracks are taken from\citet{vw94}. The
  post-RGB evolutionary track masses are estimated from the RGB
  luminosity-core mass relation obtained from fits to the evolutionary
  tracks of \citet{bertelli08}.}
\vspace{-3.0em}
\label{hr}
\end{center}
\end{figure}

The known distances to the Magellanic Clouds enabled luminosity
estimates for all the objects, which led to one of the most important
results of this survey: the unexpected discovery of a group of
evolved, dusty objects with luminosities lower than
the RGB tip luminosity ($\sim$\,2500\Lsun). We found 42 such
objects in the SMC (Paper I) and 119 such objects in the LMC (Paper II).  These objects have mid-IR
excesses and stellar parameters (other than luminosity) similar to those of post-AGB
stars. They are of A\,$-$\,K spectral types, low $\log g$
(mostly between 0 and 2 with a small number of the hotter, lower luminosity objects having log \textit{g} up to 3), and with metallicities
lower than the mean metallicity of young stars in their host galaxy. 
Because of their relatively low luminosity (100 - 2500\,\Lsun),
it is likely that these objects are dusty post-RGB
stars whose evolution was terminated by binary mass transfer or
by a merger when the star was dusty and on the RGB.  Figure~\ref{hr}
shows the positions of our dusty LMC and SMC post-RGB objects in the
HR-diagram. 

In the Galaxy, dusty single and
binary objects assumed to be post-AGB stars have been observationally well studied. However,
the unknown distances and hence luminosities do not allow for the
identification of possible dusty post-RGB stars among these
objects. Therefore, our Magellanic Cloud objects are the first examples of such
systems.
\vspace{-1.0em}
\section{Interloping objects in other evolutionary stages}
\label{interlopers}
\vspace{-0.75em}
Other classes of objects that have luminosities similar to those of
our dusty post-RGB candidates are intermediate-mass core-He burning (IM-CHeB) stars, dusty 
pre-main-sequence (PMS) stars and early-AGB (EAGB)
stars. These objects are potential interlopers contaminating our post-AGB and 
post-RGB sample. Each class of potential interlopers is considered below.
\vspace{-0.75em}
\begin{enumerate}
\item{IM-CHeB stars: The core-He burning phase is not predicted to involve heavy
mass-loss, and hence stars in this phase are not expected to have circumstellar
dust, contrary to the situation for stars in our
sample. Furthermore,  the 
metallicity distribution of our post-RGB objects peaks at a lower
value (Z $\approx$ 0.0016) than the typical metallicity of young stars in the host
galaxies (Z $\approx$ 0.004 in the SMC and 0.008 in the LMC), while the metallicty
distribution of IM-CHeB stars is expected to peak at the typical metallicity of the
young stars.  This implies that our objects 
are of lower mass (older) than IM-CHeB stars. 
In addition, our objects have relatively low $\log g$\, values.
If these stars were IM-CHeB stars, then they would be
$\sim$\,8\,$-$\,12 times more massive than
post-RGB stars at the same luminosity (see Fig.~\ref{hr}) and would therefore have a $\log g$ higher by
$\sim$1. This is not consistent with our derived $\log g$ values.  In summary,
for the above three reasons, it seems unlikely that IM-CHeB stars can explain our objects.}
\item{PMS stars: Although PMS stars have mid-IR excesses
and luminosities similar to stars in our sample, at a given luminosity, the mass of a PMS star is about 15\,$-$\,20 times that of the
corresponding post-RGB star (see Fig.~\ref{hr}), leading to a difference of $\sim$1.2 in
$\log g$ between PMS and post-RGB stars.  This gravity difference allowed us to separate the two classes
of objects (see Papers I and II for full details).  Supporting evidence for our
separation of PMS and post-RGB stars is provided by the low metallicity
of our post-RGB objects compared to the typical metallicity of young stars in the host
galaxies.
}
\item{EAGB stars: Stars
with 5\,$\ga$\,$M$\,$\ga$\,1.85\,\Msun\,can fill their Roche lobes and produce
post-EAGB stars at luminosities similar to that of our objects
(100\,\Lsun\,up to the low-mass RGB tip at $\sim$2500\,\Lsun).  
These intermediate-mass stars do not form degenerate cores or reach high luminosities on their RGB, so that they
attain larger radii and can fill their Roche lobes during their higher luminosity EAGB phase.  
It is therefore possible that there are some post-EAGB objects in our
sample. Moreover, since the thermally pulsing AGB starts below 2500\,\Lsun\,for $M$\,$\la$\,2.2\,\Msun, 
a few post-AGB stars with 2.2\,$\ga$\,$M$\,$\ga$\,1.85\,\Msun\, may be
included. We note that an EAGB star of low mass ($M$\,$\la$\,1.85\,\Msun) and with a luminosity lower that the RGB tip will have been larger
during its RGB evolution at the same or have had higher luminosities, so that
if binary interaction occurs, it will happen on the RGB and not the
EAGB. For $M$\,$\ga$\,5\,\Msun, the EAGB phase is more luminous than 2500\,\Lsun.
}
\end{enumerate}
\vspace{-2.0em}
\section{Connection to RGB binary population models}
\label{pre+pro}
\vspace{-0.75em}
\begin{figure}
\begin{center}
\resizebox{0.95\hsize}{!}{ \includegraphics{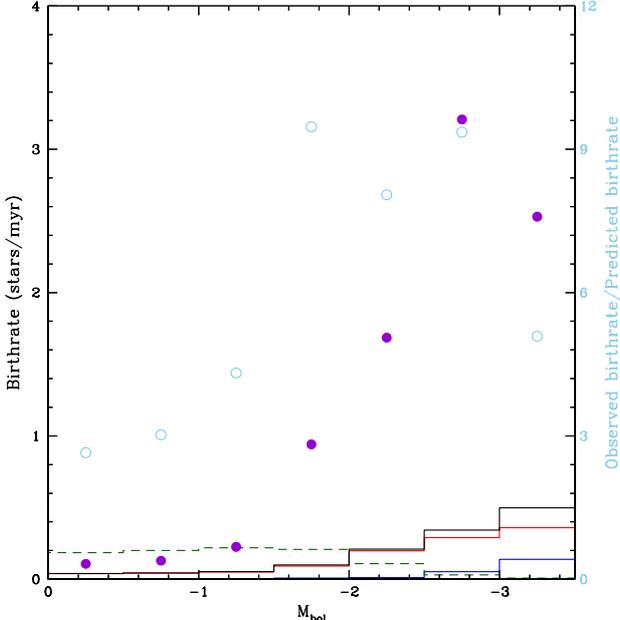}}
\vspace{-1.5em}
\caption{Luminosity distributions showing the comparison between the
total observed birthrates (stars per 1000 years) at luminosities below the RGB tip 
(post-RGB plus post-EAGB stars) and the predicted rates of
production. The purple dots are the observed birthrates,
the black solid line represents the total predicted rate of production,
which is made up of the post-RGB production rate (red line) and the post-EAGB 
production rate (blue line).  The light blue open circles show the ratio of 
the observed to the total predicted birthrate. 
The green dashed line represents the birthrate of mergers on the RGB.}
\vspace{-2.50em}
\label{theo_obv}
\end{center}
\end{figure}
To place the newly discovered dusty post-RGB systems in an
evolutionary context, it is important to understand their formation
channels and establish connections to possible precursors and
progeny. Based on their low luminosities, we expect 
that the objects reside in binary systems since single-star
mass loss that occurs during the RGB phase is insufficient to remove
the H-rich envelope and produce a dusty post-RGB star \citep[e.g.][]{vw93}. 
Therefore, we consider binary formation channels to explain their
evolution off the RGB.  

It is by now well established that variable giants form distinct
  sequences in the period-luminosity (P-L) diagram as discovered first in the LMC
  by \cite{wood99-pr}. One of the sequences, sequence-E,
  is associated with close-binary red giants that show ellipsoidal light variations 
\citep{nicholls10}. The red giants in these systems are about to fill their Roche
lobes, leading to a CE event or some less catastrophic mass transfer 
to the binary companion.  Population-synthesis models
normalised to the population of sequence-E ellipsoidal binaries in the LMC
\citep{nie12} can be used to predict the birthrate of 
post-RGB, post-EAGB stars and mergers on the RGB
relative to the number of red giant stars in the LMC. 
Here, we try to establish an evolutionary connection
between the sequence-E stars and post-RGB stars by comparing the theoretically 
predicted birthrates from \citet{nie12} with the observationally
determined birthrates of our new sample of dusty post-RGB
stars. Since the population-synthesis models of \citet{nie12} are 
normalised to the population of sequence-E binaries in
the LMC, we perform the comparison only for the LMC. 

First, we obtained predicted birthrates from the 
models of \cite{nie12}.  We used their standard model, which is normalized
to the \cite{soszynski04} sequence-E data for the LMC.  
This model of \cite{nie12} gives, in arbitrary units,  
the relative birthrates of post-RGB stars
($br{-}prgb$), post-EAGB stars ($br{-}peagb$), stellar mergers of the 
RGB ($br{-}mergers$) and stars passing through the tip
of the RGB ($br{-}tiprgb$).  \cite{nie12} found that stars take 
$2.77{\times}10^{6}$ years to evolve through the top 1 magnitude of
the RGB, so that the number of stars we expect to see at any given time in this magnitude
interval is $n{-}topmag = 2.77{\times}10^{6}\times br{-}tiprgb$.  
The actual total number of stars observed in the top 1 magnitude of the RGB
in the fields searched for post-RGB stars is 118927 (see Paper II).
Hence, the total birthrates of post-RGB and post-EAGB stars predicted by
the models in the observed fields are $br{-}prgb \times(118927/n{-}topmag)$ and
$br{-}eagb \times(118927/n{-}topmag)$, respectively.  These birthrates, in half-magnitude
bins, are shown in Fig.~\ref{theo_obv} along with their sum, which represents the
total birthrate. We note that the post-RGB birthrate dominates the post-EAGB birthrate,
although the latter increases to about 25\% of the total birthrate at the highest
luminosities.  We also show in Fig.~\ref{theo_obv} 
is the birthrate for mergers on the RGB.
Mergers are more common at lower luminosities.

Next, we estimated the total observational birthrate of post-RGB stars
in our LMC fields.  To do this, we divided the number of observed post-RGB stars
(in luminosity bins of half a magnitude) by the estimated 
evolutionary lifetime in each bin.
The evolutionary lifetime was obtained using the post-RGB evolutionary tracks of 
\citet{driebe98}.  A feature of post-RGB (and post-AGB) evolutionary tracks
is that they speed up dramatically as the envelope mass becomes low
and the star moves from the RGB to higher $T_{\rm eff}$. 
We rather arbitrarily assumed that the post-RGB phase starts at 
$\log T_{\rm eff} = \log T_{\rm eff}({\rm RGB})+0.05$ 
and continues to $\log T_{\rm eff} = 4$.
Our observed samples of LMC post-RGB stars used to calculate birthrates were
selected from the full sample using this selection criterion. This resulted in 68 out of 119 objects.  The
objects closer to the RGB (the remaining 51 objects, see Fig.~\ref{hr}) could be merged objects.
To obtain corresponding lifetimes, we extracted evolutionary 
times from the post-RGB evolutionary tracks
of \citet{driebe98} for the interval 
$\log T_{\rm eff}({\rm RGB})+0.05 < \log T_{\rm eff} < 4$.
We note that the \citet{driebe98} tracks include some mass loss,
which will cause a speed up of the evolution rate compared to the rate 
for a post-RGB star that has no external mass loss.

By dividing the observed number of stars in each luminosity bin
by the evolutionary lifetime for that bin, we derived a
partial observational birthrate.  As noted in Paper II,
our search for post-AGB and post-RGB stars is incomplete
since we took spectra of only a fraction of the possible candidates.
We found 119 post-RGB candidates, but estimated that there are another
751 post-RGB candidates in the whole LMC.  Our observed fields contain
only 56\% of the whole LMC input catalogue, therefore we expect 
0.56$\times$751\,=\,421 additional post-RGB candidates
in our observed fields.  Hence, to obtain an observational
estimate of the total post-RGB birthrates for our observed fields, we need to multiply each 
partial observational birthrate by a factor of
(421+119)/119\,=\,4.5. These final observed post-RGB birthrates are shown as large
solid points in Fig.~\ref{theo_obv}.

It is clear from Fig.~\ref{theo_obv} that the observationally estimated
post-RGB birthrate is higher than the theoretically predicted
birthrate, especially at higher luminosities. We find that the average ratio of observed to predicted birthrate is 6.0.
There are various possibilities that could explain this discrepancy,
and we discuss each of these below.
\vspace{-0.75em}
\begin{enumerate}
\item{An over-estimation of the incompleteness factor: 
Using results from Paper II, we estimated
that our observed post-RGB sample was incomplete by the large factor of 4.5. The uncertainty in this completeness
factor could explain a significant part of the discrepancy. }
\item{An underestimate of the post-RGB evolution time: The \citet{driebe98} evolutionary tracks include mass loss given by
a \citet{reimers75}-type law which, in the middle of the luminosity
range considered here ($\log L/{\rm L}_{\odot} = 2.6$) 
and at $\log T_{\rm eff}\,=\,3.75$, is essentially identical to the
rate of consumption of the hydrogen-rich envelope by nuclear burning.
Since the evolution rate is proportional to the rate of reduction
of the envelope mass by the combination of 
mass loss and nuclear burning, our post-RGB
lifetimes could be too short by a factor of $\sim$2 if post-RGB stars 
have no mass loss.  If they have mass re-accretion from a circumbinary
disk, which is often observed in Galactic post-AGB binaries \citep[e.g.][]{vanwinckel03},
then our assumed post-RGB lifetimes could be shorter by an even 
larger factor, and our observational birthrates will be over-estimated
by the same factor.}
\item{Uncertainties in the model post-RGB birthrate: 
The population of sequence-E stars in the LMC is well determined by OGLE monitoring \citep{soszynski04}.
These stars are about to fill their Roche lobes, and their evolution
up the RGB on nuclear timescales is relatively simple.  For this reason we 
chose these stars and the modelling of \citet{nie12} as a way to derive
a reasonably accurate birthrate estimate for post-RGB stars.  We considered
all stars that fill their Roche lobes on the RGB as the precursors of our
post-RGB stars.
In practice, it is likely that only a fraction of the Roche-lobe-filling stars
are the precursors of post-RGB stars
because it is possible that the stars that undergo CE
evolution on filling their Roche lobes will transit rapidly
to $\log T_{\rm eff} > 4$ and will not form part of our observational sample.
Thus our birthrates estimated from the models are likely over-estimates
and not the under-estimates required if we are to match the observed birthrates.
Modelling the uncertainties can therefore probably not explain the 
discrepancy between the observed and predicted post-RGB star
birthrates. We note that although we have only mentioned post-RGB stars in this discussion,
we implicitly included the post-EAGB component of the
population.}
\item{Contamination of our post-RGB sample by interlopers 
especially PMS stars. As discussed in Sect. 3,
contamination is probably not a significant problem.}
\item{Our choice of $\log T_{\rm eff}({\rm RGB})+0.05$ as the lower limit for $\log T_{\rm eff}$
of post-RGB stars.  We tested this by repeating our calculations using $\log T_{\rm eff}({\rm RGB})+0.1$ as
the lower limit for $\log T_{\rm eff}$ and found an average ratio of observed to predicted birthrate 
of 7.3 instead of 6.0.  Therefore, our choice of the lower limit for $\log T_{\rm eff}$ does not
greatly influence the discrepancy.}
\end{enumerate} 
\vspace{-1.25em}
A related possible problem for our post-RGB 
objects, at least at the lower luminosities, is their expected
lifetime as they evolve from the RGB to $\log T_{\rm eff} = 4$.
According to the tracks of \citet{driebe98}, this lifetime is $\sim$$10^3$ years
for the highest luminosity RGB stars and nearly $10^8$ years
for the lowest luminosity objects in our sample.  The possible problem with
this lifetime is the requirement that dust remain in the vicinity of the
post-RGB stars while they evolve to $\log T_{\rm eff} = 4$.  However, we
note that in a typical post-AGB binary, the dust is in a circumbinary disk \citep{vanwinckel09}
with a long potential lifetime, and we expect a similar situation to apply for
post-RGB stars.  Until studies of the lifetimes of such disks are made,
we cannot conclude whether the long evolutionary lifetimes are a problem.

In light of the above discussion, we think the discrepancy between the
observed post-RGB birthrate estimate and the model birthrate estimate
is most likely due to the uncertainty in our incompleteness factor and
to uncertainties in the post-RGB evolution rate.  It is highly desirable 
that a larger sample of mid-infrared-selected post-RGB candidates be
examined spectroscopically to improve the incompleteness factor for
post-RGB stars in the LMC.  It is also desirable that more post-RGB and
post-EAGB evolutionary tracks be made with modern input physics and
for a range of mass loss and mass re-accretion rates to better
determine the post-RGB evolution timescale. 

Finally, we note that low on the RGB in Fig.~\ref{theo_obv}, the
birthrate of mergers exceeds the birthrate of binary post-RGB stars.
Such stars are likely to eject part of their envelope during the
merger process, and they might be expected to show a mid-IR excess for
some time.  Our post-RGB sample was selected because of a mid-IR
excess, so it is possible that those stars in our sample that are close to
the RGB in \teff\, (see Fig.~\ref{theo_obv}) have recently undergone a merger.
\vspace{-1.25em}
\section{Conclusions} 
\label{results}
\vspace{-0.75em}
We presented a new population of evolved,
dusty, low-luminosity, low-metallicity, low $\log g$ post-RGB stars in the Magellanic Clouds.  
These objects are
very similar to post-AGB binaries, except that they have lower 
luminosities (and hence masses), and they are very likely to have evolved off the RGB via binary
interaction. We compared the observed birthrate of the post-RGB sample to
the predictions from population-synthesis models of \citet{nie12}, 
which were normalised to the LMC population of sequence-E ellipsoidal  
binaries on the RGB. 

We found that there is order-of-magnitude agreement between the observed and 
predicted birthrates of post-RGB stars given the uncertainties involved, the most important uncertainty being the incompleteness
  factor in the observed sample of post-RGB stars and our poor knowledge of
post-RGB evolution rates. We therefore conclude that our dusty post-RGB
  stars are very likely to represent a newly discovered phase of binary stellar
  evolution.   We also note that mergers are relatively common 
low on the RGB and that stars low on the RGB with 
mid-IR excesses may have recently undergone a merger.
\vspace{-0.5em}
\begin{acknowledgements}
\vspace{-0.5em}
DK and HVW acknowledge the support of the KU Leuven contract
GOA/13/012. DK acknowledges the support of the FWO grant G.OB86.13.
PRW has received support from the Australian Research Council Discovery Project DP120103337.
\vspace{-1.5em}
\end{acknowledgements}
\vspace{-1.5em}

\bibliography{mnemonic,devlib}

\begin{thebibliography}{21}
\expandafter\ifx\csname natexlab\endcsname\relax\def\natexlab#1{#1}\fi

\bibitem[{{Bertelli} {et~al.}(2008){Bertelli}, {Girardi}, {Marigo}, \&
  {Nasi}}]{bertelli08}
{Bertelli}, G., {Girardi}, L., {Marigo}, P., \& {Nasi}, E. 2008, A\&A, 484, 815

\bibitem[{{Bertelli} {et~al.}(2009){Bertelli}, {Nasi}, {Girardi}, \&
  {Marigo}}]{bertelli09}
{Bertelli}, G., {Nasi}, E., {Girardi}, L., \& {Marigo}, P. 2009, A\&A, 508, 355

\bibitem[{{Driebe} {et~al.}(1998){Driebe}, {Schoenberner}, {Bloecker}, \&
  {Herwig}}]{driebe98}
{Driebe}, T., {Schoenberner}, D., {Bloecker}, T., \& {Herwig}, F. 1998, A\&A,
  339, 123

\bibitem[{{Han} {et~al.}(2002){Han}, {Podsiadlowski}, {Maxted}, {Marsh}, \&
  {Ivanova}}]{han02}
{Han}, Z., {Podsiadlowski}, P., {Maxted}, P.~F.~L., {Marsh}, T.~R., \&
  {Ivanova}, N. 2002, MNRAS, 336, 449

\bibitem[{{Heber}(2009)}]{heber09}
{Heber}, U. 2009, ARA\&A, 47, 211

\bibitem[{{Ivanova} {et~al.}(2013){Ivanova}, {Justham}, {Chen}, {De Marco},
  {Fryer}, {Gaburov}, {Ge}, {Glebbeek}, {Han}, {Li}, {Lu}, {Marsh},
  {Podsiadlowski}, {Potter}, {Soker}, {Taam}, {Tauris}, {van den Heuvel}, \&
  {Webbink}}]{ivanova13}
{Ivanova}, N., {Justham}, S., {Chen}, X., {et~al.} 2013, A\&ARv, 21, 59

\bibitem[{{Kamath} {et~al.}(2014){Kamath}, {Wood}, \& {Van Winckel}}]{kamath14}
{Kamath}, D., {Wood}, P.~R., \& {Van Winckel}, H. 2014, MNRAS, 439, 2211

\bibitem[{{Kamath} {et~al.}(2015){Kamath}, {Wood}, \& {Van Winckel}}]{kamath15}
{Kamath}, D., {Wood}, P.~R., \& {Van Winckel}, H. 2015, ArXiv e-prints

\bibitem[{{Moe} \& {De Marco}(2006)}]{moe06}
{Moe}, M. \& {De Marco}, O. 2006, ApJ, 650, 916

\bibitem[{{Nicholls} {et~al.}(2010){Nicholls}, {Wood}, \& {Cioni}}]{nicholls10}
{Nicholls}, C.~P., {Wood}, P.~R., \& {Cioni}, M.-R.~L. 2010, MNRAS, 405, 1770

\bibitem[{{Nie} {et~al.}(2012){Nie}, {Wood}, \& {Nicholls}}]{nie12}
{Nie}, J.~D., {Wood}, P.~R., \& {Nicholls}, C.~P. 2012, MNRAS, 423, 2764

\bibitem[{{Reimers}(1975)}]{reimers75}
{Reimers}, D. 1975, {Circumstellar envelopes and mass loss of red giant stars}
  (Problems in stellar atmospheres and envelopes.), 229--256

\bibitem[{{Sch{\"o}nberner}(1983)}]{schoenberner83}
{Sch{\"o}nberner}, D. 1983, ApJ, 272, 708

\bibitem[{{Soszynski} {et~al.}(2004){Soszynski}, {Udalski}, {Kubiak},
  {Szymanski}, {Pietrzynski}, {Zebrun}, {Szewczyk}, {Wyrzykowski}, \&
  {Dziembowski}}]{soszynski04}
{Soszynski}, I., {Udalski}, A., {Kubiak}, M., {et~al.} 2004, Acta Astronomica,
  54, 347

\bibitem[{{Tognelli} {et~al.}(2011){Tognelli}, {Prada Moroni}, \&
  {Degl'Innocenti}}]{tognelli11}
{Tognelli}, E., {Prada Moroni}, P.~G., \& {Degl'Innocenti}, S. 2011, A\&A, 533,
  A109

\bibitem[{{Van Winckel}(2003)}]{vanwinckel03}
{Van Winckel}, H. 2003, ARA\&A, 41, 391

\bibitem[{{Van Winckel} {et~al.}(2009){Van Winckel}, {Lloyd Evans}, {Briquet},
  {De Cat}, {Degroote}, {De Meester}, {De Ridder}, {Deroo}, {Desmet},
  {Drummond}, {Eyer}, {Groenewegen}, {Kolenberg}, {Kilkenny}, {Ladjal},
  {Lefever}, {Maas}, {Marang}, {Martinez}, {{\O}stensen}, {Raskin}, {Reyniers},
  {Royer}, {Saesen}, {Uytterhoeven}, {Vanautgaerden}, {Vandenbussche}, {van
  Wyk}, {Vu{\v c}kovi{\'c}}, {Waelkens}, \& {Zima}}]{vanwinckel09}
{Van Winckel}, H., {Lloyd Evans}, T., {Briquet}, M., {et~al.} 2009, A\&A, 505,
  1221

\bibitem[{{Vassiliadis} \& {Wood}(1993)}]{vw93}
{Vassiliadis}, E. \& {Wood}, P.~R. 1993, ApJ, 413, 641

\bibitem[{{Vassiliadis} \& {Wood}(1994)}]{vw94}
{Vassiliadis}, E. \& {Wood}, P.~R. 1994, ApJS, 92, 125

\bibitem[{{Webbink}(1984)}]{webbink84}
{Webbink}, R.~F. 1984, ApJ, 277, 355

\bibitem[{{Wood} {et~al.}(1999){Wood}, {Alcock}, {Allsman}, {Alves}, {Axelrod},
  {Becker}, {Bennett}, {Cook}, {Drake}, {Freeman}, {Griest}, {King}, {Lehner},
  {Marshall}, {Minniti}, {Peterson}, {Pratt}, {Quinn}, {Stubbs}, {Sutherland},
  {Tomaney}, {Vandehei}, \& {Welch}}]{wood99-pr}
{Wood}, P.~R., {Alcock}, C., {Allsman}, R.~A., {et~al.} 1999, in IAU Symposium,
  Vol. 191, Asymptotic Giant Branch Stars, ed. T.~{Le Bertre}, A.~{Lebre}, \&
  C.~{Waelkens}, 151

\end{thebibliography}

\end{document}